\documentclass[conference]{IEEEtran}
\usepackage[english]{babel}

\usepackage{amsmath}
\usepackage{graphicx}
\usepackage{tikz}
\usetikzlibrary{arrows}
\usepackage{subfigure}

\usepackage{pgfplots} 
\usepackage[colorlinks=true, allcolors=blue]{hyperref}

\newcommand{\proposaltechnion}[1]{{}}

\newcommand{\notinconf}[1]{{}}

\title{Allowing Blockchain Loans with Low Collateral}
\proposaltechnion{
\title{Research Proposal: A way to lower collateralize loans in Blockchain}
\author{Tom Azoulay\\
The Henry and Marilyn Taub Faculty of Computer Science,\\
Technion — Israel Institute of Technology\\
Advisor: Dr. Ori Rottenstreich}
}

\begin{document}

\author{
\IEEEauthorblockN{
Tom Azoulay
}
\IEEEauthorblockA{
Technion - Israel Institute of Technology
}

\and

\IEEEauthorblockN{
Uri Carl
}
\IEEEauthorblockA{
Blue Frontiers Partners, LLC
}

\and

\IEEEauthorblockN{
Ori Rottenstreich
}
\IEEEauthorblockA{
Technion - Israel Institute of Technology
}
}
\IEEEoverridecommandlockouts
\IEEEpubid{\makebox[\columnwidth]{978-8-3503-1019-1/23/\$31.00~\copyright2023 IEEE \hfill} \hspace{\columnsep}\makebox[\columnwidth]{ }}
\maketitle
\IEEEpubidadjcol
\thispagestyle{empty}
\pagestyle{plain}
\begin{abstract}
Collateral is an item of value serving as security for the repayment of a loan. In blockchain-based loans, cryptocurrencies serve as the collateral. The high volatility of cryptocurrencies implies a serious barrier of entry with a common practice that collateral values equal multiple times the value of the loan. As assets serving as collateral are locked, this requirement prevents many candidates from obtaining loans. In this paper, we aim to make loans more accessible by offering loans with lower collateral, while keeping the risk for lenders bound. We propose a credit score based on data recovered from the blockchain to predict how likely a potential borrower is to repay a loan. Our protocol does not risk the initial amount granted by liquidity providers, but only risks part of the interest yield gained from the borrower by the protocol in the past.
\end{abstract}

\section {Introduction}
\label{Introduction}
Decentralized Finance (DeFi) is an emerging  technology for various financial services that run as smart contracts on decentralized cryptocurrencies. Smart contracts are applications stored on the blockchain that execute a code accessible to all.  
The market size of the DeFi market\footnote{DeFiLlama -  https://defillama.com/} is estimated to be over 180 billion USD in late 2021 and 39.6 billion USD in December 2022. The common DeFi applications include decentralized exchanges (DEXs) that allow the exchange of cryptocurrencies, systems enabling locking coins for earning interest, lending systems where cryptocurrencies serve as collateral, and even insurance services~\cite{DeFiAnalysis, dos2022new, EvolutionLending, insurance}.

Lending is among the most important financial activities, coming in various forms, such as government bonds, corporate debt, mortgages, student loans, and consumer loans. 
Accordingly, DeFi loans have emerged in recent years with the appearance of loan decentralized protocols, such as Aave, Compound, Cream finance, and DaoMaker/Oasis~\cite{EvolutionLending, GudgeonW0K20, SokDeFi}. Table~\ref{tab:protocols} summarizes the updated total value locked in such platforms, together with their actual loan values.

DeFi lending services include two types of loans, with a major distinction with respect to their time period length. First, a flash loan is a type of loan where a value is borrowed and returned within a single transaction without collateral~\cite{wang2020towards, FlashLoanAttacking}. Such a loan is often used to take advantage of arbitrage among DEXs, is restricted in its time period, and is often considered risky.   
On the other hand, the second type of loan, a collateralized loan, allows for longer return time periods, and is typically associated with interest rates, which are based on the loan period~\cite{GudgeonW0K20, yaish2022blockchain}. In addition to smart contract exploits~\cite{babel2021clockwork}, loans have an inherent risk of potential loss. As in traditional loans,  in DeFi lending services, collateral serves as a security in case a borrower either cannot or does not intend to return the loan.
There are different liquidations processes across the different platforms offering loans, and they mainly favor liquidators over borrowers~\cite{qin2021empirical}. There have been some proposed solutions for making DeFi lending more robust \cite{chiu2022inherent}.

In DeFi lending services, the loan is often given in stablecoins, whose values tightly follow the USD. On the other hand, collateral is given in cryptocurrencies of large market cap, like ETH, the native coin of the Ethereum network, as well as in other cryptocurrencies of smaller market cap.
In addition to the regular risk inherent in one's ability to return loans, DeFi lending has another source of risk, namely, the volatility of cryptocurrencies. This additional risk is mitigated by the requirement for over-collateralization - a requirement that the value of the collateral is higher than the loan value. As the ratio of collateral to loan value changes constantly, based on market prices, to avoid  under-collateralization, lending protocols often define minimal thresholds (with typical values of 120-300\%) for various cryptocurrencies, such that when the collateral to loan ratio falls below the threshold, the loan is liquidated and the collateral is sold. This eliminates the need to return the loan. To reduce the chances of the collateral becoming liquidated, in practice, a borrower often provides collateral of even higher ratios. This implies low loan accessibility with high entrance bars for potential borrowers. 

 


In this study, we propose a new way to make loans more accessible by requiring less collateral. This protocol will allow offering users a less collateralized loan, widening the number of people who can get it. The reduction is provided based on a mix of their past contribution to the protocol and a risk assessment (a crypto-based credit score). This allows lowering each time the collateral or to contract a bigger loan with the same collateral. The proposal presents a risk on the lender's gains, but not their initial staking, and the credit score mitigates this risk.

For credit scoring, the literature is vast and entertains many binary classification models. See~\cite{creditScoreLit} for a good survey. Machine learning algorithms are often used, but the baseline model is typically a logistic or probit model, as they are the simplest to implement and interpret. We, therefore, start with a Probit model here. Moreover, for small datasets, overly sophisticated machine learning models can overfit the data. A basic Probit model can be advantageous in that regard. Finally, credit scoring on blockchain platforms, in particular, has some advantages over traditional banking platforms, including timeliness and accuracy of loan-related data \cite{bystrom2019blockchains}.

\begin{table}[t!]
\centering
\resizebox{\linewidth}{!}{
\begin{tabular}{ c|cc }
\textbf{Platform}  & Total Value Locked  & Loan Values \\
&  (TVL, in USD)  & (in USD)   \\
\hline
Aave~\cite{AaveWatch}  &   4B  & 1.5B        \\
\hline
Compound~\cite{compound} &    2B  &    600M    \\
\hline
Cream~\cite{DefiLama} &   42M &  1.3M  \\ 
\hline
MakerDAO/Oasis~\cite{MakerOasisStats} & 2B  & NA     \\ 
\end{tabular}
}
\caption{Major Liquidity Market Platforms allowing Blockchain-based Loan Services }
\label{tab:protocols}
\end{table}

\begin{figure*}[t!]
\centering
\begin{tikzpicture} 
\begin{axis} 
[  
    ybar, 
    legend style={at={(0.4,0.85)}, 
     anchor=north,legend columns=-1},   
    height = 0.33 \textwidth,
    width = 0.92  \textwidth, 
    ylabel={Threshold Liquidation (\%)}, 
    symbolic x coords={ETH, WBTC,DAI+TUSD+USDC,  AAVE, UNI,BAT, SNX},  
    xtick=data,
    tick label style={font=\footnotesize},
    every node near coord/.append style={font=\footnotesize},
    nodes near coords,  
    nodes near coords align={vertical},  
    ]  
\addplot coordinates { (ETH, 125) (DAI+TUSD+USDC, 125) (WBTC, 153.8) (UNI, 153.8) (AAVE, 153.8) (SNX, 250) (BAT, 153.8)};
\addplot coordinates {(DAI+TUSD+USDC, 119) (ETH, 121.95) (UNI, 133.3) (WBTC, 142.9) (AAVE, 137) (BAT, 153.8)};  
\addplot coordinates {(DAI+TUSD+USDC, 133) (ETH, 133) (UNI, 153.8) (WBTC, 133) (AAVE, 166.7) (SNX, 222)};  
\addplot coordinates { (ETH, 130) (WBTC, 130) };

\legend{Aave, Compound, Cream, MakerDAO/Oasis}

\end{axis}  
\end{tikzpicture}  
\caption{Threshold of liquidation based on the type of collateral for different lending platforms.}
\label{ThredLiq}
\end{figure*}
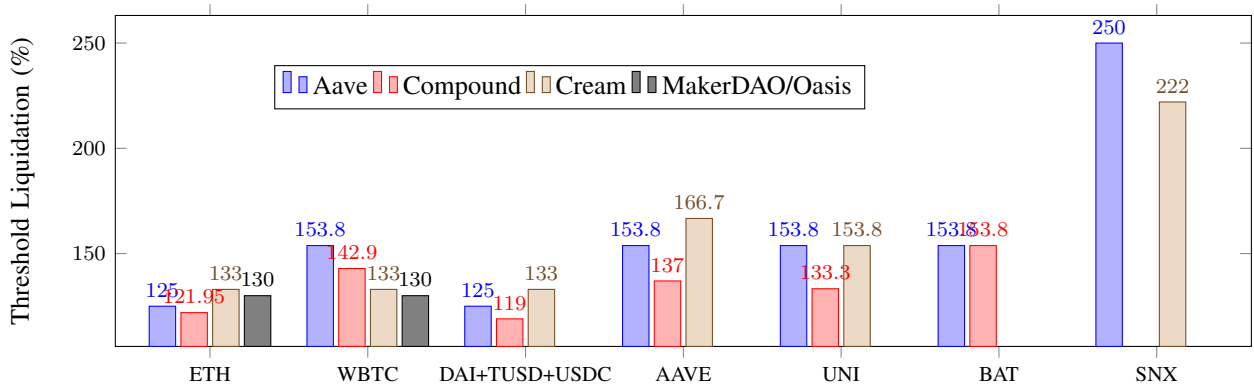
\emph{Contributions and paper overview.} 
In Section~\ref{section_background} we overview the basic mechanism of DeFi lending protocols and present the terminology of the paper. Next, in Section~\ref{section_related_work} we survey existing protocols and related literature. Section~\ref{section_protocol_approach} presents our proposed protocol that allows collateral of lower values.
As part of the protocol design, we propose in Section~\ref{section_risk} a new model to estimate the risk of borrowers based on their account history and interactions with the protocol. We study data of a lending protocol named Compound in Section~\ref{section_compound_data} and analyze it to tune our protocol. We conduct experiments to evaluate the advantages of the proposed protocol in Section~\ref{section_evaluation}. Finally, conclusions and directions for future work can be found in Section~\ref{section_conclusions}.

\section{Background on DeFi Lending Protocols}
\label{section_background}
\textbf{Glossary.} A loan is an amount of money that is borrowed and has to be paid back, usually together with an extra amount of money called interest.  Lending in the world of blockchain concerns the lending of cryptocurrencies  (it may be one or multiple) and is enabled by smart contracts. 
We focus on long-term loans in which collateral should be provided by the borrower. 

We use the following common terminology:

\emph{protocol}: A smart contract allowing interaction of the various actors to supply collateral, take loans, or deposit liquidity. 

\emph{borrower}: The entity receiving the loan, often in stablecoins, like USDC.

\emph{collateral}: The guarantee provided by the borrower, usually in cryptocurrency. 

\emph{lender}: An entity providing liquidity to the protocol in exchange for an interest rate. A lender does not interact with a particular borrower.


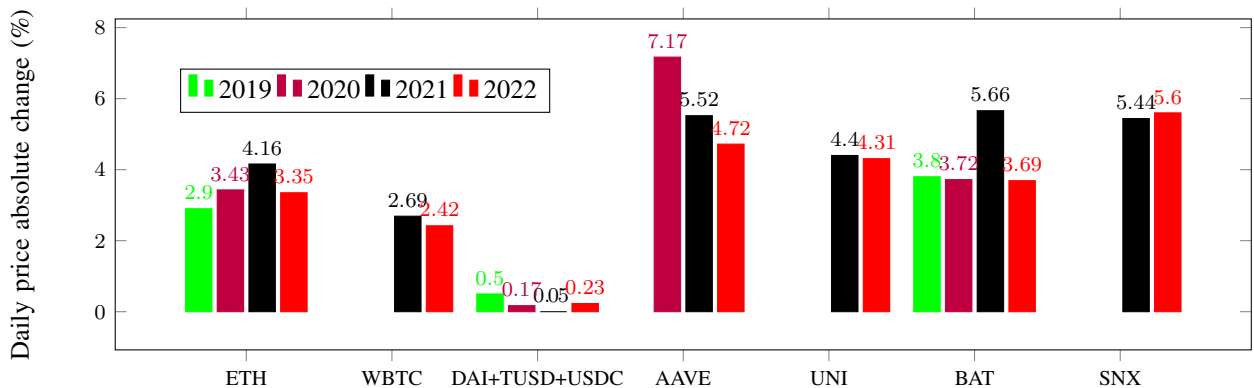
\begin{figure*}[t!]
\centering
\begin{tikzpicture}  
\begin{axis}  
[  
 ybar, 
  legend style={at={(0.22,0.85)}, 
      anchor=north,legend columns=-1},     
    ylabel={Daily price absolute change (\%)}, 
    enlargelimits=0.15,
    height = 0.33 \textwidth,
    width = 0.92 \textwidth,
    tick label style={font=\footnotesize},
    symbolic x coords = {ETH, WBTC, DAI+TUSD+USDC, AAVE, UNI,BAT, SNX},
    every node near coord/.append style={font=\footnotesize},
    xtick={ETH, WBTC, DAI+TUSD+USDC, AAVE, UNI,BAT, SNX}, 
    nodes near coords,
    nodes near coords align={vertical},  
    ]  
\addplot [green,fill=green] coordinates {(ETH, 2.9) (DAI+TUSD+USDC, 0.5)  (BAT, 3.8)};
\addplot [purple,fill=purple] coordinates {(ETH, 3.43) (DAI+TUSD+USDC, 0.17)  (AAVE, 7.17) (BAT, 3.72)};  
\addplot [black,fill=black] coordinates {(ETH, 4.16) (WBTC, 2.69) (DAI+TUSD+USDC, 0.0)  
(UNI, 4.4)  (AAVE, 5.52) (SNX, 5.44) (BAT, 5.66)};  
\addplot [red,fill=red] coordinates {(ETH, 3.35) (WBTC, 2.42) (DAI+TUSD+USDC, 0.23)  (UNI, 4.31)  (AAVE, 4.72) (SNX, 5.60) (BAT, 3.69)};
\legend{2019,2020,2021,2022} 
\node[font=\footnotesize, anchor=west, right] at (207.5,46) {$5$};
\node[font=\footnotesize, anchor=west, right] at (190.5,46) {$0.$};
\end{axis}  
\end{tikzpicture}  
\caption{Mean of daily price change in \% for multiple currencies across different years (Data used from Invest)}
\label{VolYearFromInvest}
\end{figure*}

\textbf{Interactions with the Protocol. }
 A loan with collateral works in the following way.
 An account (user) first supplies some collateral to the protocol for the right to take loans. The account can then ask for a loan of some value. The collateral is then locked until the debt (including interest) is repaid. For each type of collateral (a particular cryptocurrency), the protocol defines a liquidation threshold that serves as a lower ratio between the collateral value and the loan value. 
 While the lower bound is satisfied, the exact amount of the ratio is due to the choice of the account. When the value of the collateral drops and reduces the ratio below the threshold, the collateral is sold by the protocol instead of the payback of the loan.
 
 Providing collateral of values higher than the loan amount is required for several reasons. First, it aims to motivate the borrower to return the loan. Second, it deals with the risk of high volatility in the value of the collateral. Because of network congestion between the decision and the liquidation process by the network, there might be a delay (as for every transaction). During this delay, the value of the collateral may drop even further. To sum up, this type of loan can end in one of  three possible ways:

\emph{Case 1}: The borrower repays the loan fully with interest. 
The collateral is unlocked and returned to the borrower.

\emph{Case 2}: The borrower fails to pay for a due date of repayment, thereby defaulting. The protocol liquidates the collateral and the loan is terminated. 
 
\emph{Case 3}: The borrower's collateral passes under the liquidation threshold due to the price volatility of the assets provided. The protocol considers the borrower to have defaulted, and liquidates the collateral, in turn terminating the loan.

We provide an example for the third case. Consider two users borrowing a loan of amount X. Assume they both provide the same type of collateral and that the liquidation threshold for this type of collateral is 130\%. Meaning, if the value of the collateral is at 130\% the value of the loan or under, the collateral is automatically liquidated by the protocol. User 1 provides a collateral of value 2.5X and User 2 provides one of value 3X. There is a price change causing the collateral to lose half of its value. User 1's loan is liquidated, as it has passed under the liquidation threshold, and User 2's loan is not liquidated. User 1 could have prevented liquidation by providing further collateral.

 
We focus on long term loans (rather than flash loans) in order to assess the risk of a user defaulting, to allow for better decision-making before approving the loan and to provide greater access to the lending system by lowering the required collateral. As flash loans are already accessible without the need for high funds, our study does not focus on them.
 

\textbf{Lending Platforms Statistics.} We collected information from the four major DeFi lending platforms Aave, Compound, Cream, and MakerDAO/Oasis, previously mentioned in Table~\ref{tab:protocols}. In each platform,  ratios are presented for some of the dominant cryptocurrencies. 

Figure~\ref{ThredLiq} shows the minimal collateralization ratio  by each platform. For instance, in the Compound platform, in case the collateral is in ETH, its value should be at least 121.95\% of the value of the loan, and of 142.9\%, in the case of collateral in WBTC.  In the case the collateral drops below this ratio, the loan is automatically liquidated. The ratio required depends on the type of assets that serve as collateral. The graphs were constructed using data from the lending protocols' platforms (as of December 2022, susceptible to change over time).

Figure~\ref{VolYearFromInvest} shows the mean of the daily change in price in \% of different currencies across different years. The change in \% is computed using the closure price of the previous day, and then computing the \% change. Since AAVE replaced the LEND token in 2020, we exclude this effect and start only on the value from Nov 6 2020 (compared with Nov 5 and so on).

\section{Related Work}
\label{section_related_work}

\subsection{Traditional Credit Scoring} 
Credit score, as defined by the oxford dictionary, is a number assigned to a person that indicates to lenders the person's capacity to repay a loan~\cite{CreditScoreOxford}.  
Typically, the process to calculate a credit score is by first modeling one's probability of default for a given loan over some future horizon. The probability, between 0 and 1, is then scaled to arrive at a score on a larger range of integers. The FICO score, for example, has a range of 300 to 850.

To calculate the probability of default, traditional approaches focus on binary classification models, where the question is if one will default (1) or not (0), say, over a 3-month horizon. Broadly speaking, these models fall into one of two categories: ensemble methods and individual classification methods. Ensemble methods aggregate multiple models, such as random forests and neural networks. Individual classification methods focus on one model, be it linear, like logistic regression, or non-linear, like Naive Bayes.s~\cite{creditScoreLit} lays out multiple machine-learning approaches.

The question of which features to include in the model is another nuance in these models. For consumer loans in particular, some data can be obtained from an external credit bureau, while other data particular to the account and loan can be extracted from information provided by the specific individual. For legal reasons, some identifying attributes have to be masked. See~\cite{KHANDANI20102767} for a detailed approach of useful consumer data for this problem, and the data omitted due to legal restrictions.

This paper borrows approaches from the traditional credit history literature but amends the features and models, based on the availability of data, and the differences of the blockchain lending platform from traditional lending processes.

\subsection{Existing commercial platforms} 
Companies of the first type mentioned in the \ref{Introduction} provide loans based on collateral to ensure being reimbursed in case of a default by the client. The threshold serves as a volatility countermeasure. Because some time may pass between the liquidation order and the processing of it, due to blockchain network latency, the price may further drop. Thus the over-collateralization of the asset comes to counter this negative effect and ensure full repayment of the loan. Theoretically, this type of loan may lose money if, during the time of processing the liquidation, there is a very high volatility, causing the collateral to drop under 100\% of the value of the loan.

Companies of the second type, which we will now present, provide loans based on mixed insurance. The collateral and the credit assessment are made for a client. There exists a risk for a default that is mitigated in theory by other truthful borrowers. These companies gain from attracting more customers even if they expose themselves to larger risks.

There are multiple studies tackling this risk problem currently, but to they often do not make their tools, full methodology, and results available for comparison.


Rocifi~\cite{RociFi} mentions scoring people based on "dozens of data points across several EVM-compatible chains: borrowing and repayment behavior on DeFi lending protocols, DAO contributions, liquidity provision, and trading activities, balance changes over time, etc". The full scoring model and performance, however, are not public.

Arcx~\cite{ARCx} is also based on "historical on-chain borrowing activity". Its documentation mentions it rewards users on the past 120 days and favors those borrowing in the middle range, with 60\% mentioned as the optimum borrowing rate. The second parameter is based on all past history and how close one comes to being liquidated. Finally, the last parameter is a penalty of liquidation. Though the method is relatively detailed, the results of those estimations are not made public. 

Creda~\cite{CreDA} bases its score on a mix of on-chain and off-chain data. The use of off-chain data prohibits privacy, contrary to our approach.

Trava~\cite{TRAVA} uses data from multiple networks based mainly on on-chain data from Binance Smart Chain transactions, including the native transferring transactions and the transactions to deposit, withdraw, repay, and borrow tokens generated from Binance Smart Chain lending DApps. Use includes the age of the address, its transaction amount, its frequency of transaction, its number of liquidations, and its total value of liquidations

Quadrata~\cite{Quadrata} offers a credit score assessment but does not provide details on how credit scoring is made. 

Credefi~\cite{Credefi} does not provide any data on its the method, and relies on a licensed financial institution to serve as collateral and liquidator in all legal manners.

Spectral~\cite{Spectral} uses on-chain transaction data. The website specifically says "This data is grouped into five categories: payment history, liquidation history, amounts owed and repaid, credit mix, and length of credit history." 

Truefi~\cite{TrueFi} relies on a non-detailed human process.

Telefy~\cite{Telefy} uses the number of transactions, wallet balance, time of retaining coins, Telefy usage, NFT transactions, the amount owed and repaid, credit history length, previous liquidation(s), and cross-chain validation

Masa~\cite{Masa} uses a variety of parameters with no further details on methodology. It uses Credit Bureau data, bank transaction data, mobile money Data, on-chain data, and centralized exchange Data.

\begin{table}[t!]
\centering
\begin{tabular}{ c|cc }
\textbf{Platform} & Total Value Locked (TVL) \$   \\
\hline
Credefi~\cite{Credefi},~\cite{DefiLama} &  3M \\ 
\hline
Rocifi~\cite{RociFi},~\cite{DefiLama} & 78k    \\
\hline
Trava~\cite{TRAVA},~\cite{DefiLama} &  277k  \\ 
\hline
Truefi~\cite{TrueFi},~\cite{DefiLama} & 30.2M \\ 
\hline
Arcx~\cite{ARCx}   &   Beta Phase  \\
\hline
Creda~\cite{CreDA}  &  Beta Phase \\ 
\hline
Quadrata~\cite{Quadrata}  & Just credit Score  \\ 
\hline
Spectral~\cite{Spectral}  & Just credit Score  \\ 
\hline
Telefy~\cite{Telefy} &  No lending available yet\\ 
  \hline
Masa~\cite{Masa}  &  Goerli testnet \\ 
\end{tabular}
\caption{Liquidity Market Platforms providing credit score for Blockchain-based Loan Services}
\label{tab:topologies}
\end{table}

\subsection{Credit score solutions} 
\cite{CSLD} presents a classification of clients using limited information, like loan history, and excluding demographic or personal features. They manage to obtain a 76\% accuracy but fail to reduce the type II error (loans that should not have been granted but were) below 10\%.

To the best of our knowledge, the only research paper mentioning on-chain data-based credit scoring is \cite{Score_Aave_Creditworthiness}, aiming to build a credit score based on Aave's account history. 

The following paper \cite{privatecredit} offers a solution to user privacy using Zero Knowledge proofs and credit score calculation (CSC) on the blockchain. This research aims at performing a CSC solely by using the information provided by financial institutions while the user's identity is preserved.

Aside from \cite{Score_Aave_Creditworthiness}, the approaches mentioned either omit blockchain data, mix data (like identity, which requires external input), or do omit their performance or methodology. We believe our approach to be safer than that of ~\cite{Score_Aave_Creditworthiness} because in ours, a malicious user will always lose more than what he earns, while their paper creates an opportunity for a malicious user to game the score, as he can gain money from it.

\section{The Proposed Protocol for Low Required Collateral}
\label{section_protocol_approach}

\subsection{Protocol approach}
In a traditional loan, a user obtains a loan at a certain rate and provides in exchange the collateral that is locked until repayment. The user may come a second time after repaying his loan to obtain a new one. Our approach suggests modifying the traditional process at the second step (when the user comes for a new loan). Instead of providing him an additional loan similar to the preceding one, we offer to reduce the collateral required making a more affordable loan. The collateral required is lowered by a part of the amount gained in the past by the protocol. If the past gains are of \$X, the collateral required is lowered by a fraction of this amount. To determine what fraction of the \$X should be lowered, an estimate of how reliable the account is is calculated i.e the credit score. The amount by which the collateral is lowered represents the risk introduced in this protocol. In the case the user does not repay, some of the original gains might be lost. We present in the next subsection the mathematical model to ensure that we realize a bigger profit than a deposit at some no-risk investment opportunity that we call the bank.

Our protocol allows offering loans to a wider spectrum, as our collateral requirement is lower each time for a loan of the same amount, and therefore requires locking fewer assets. The use of only past earnings serves as a deterrent for malicious users, as the amount of the theft is always smaller than the gain of the protocol, even if the credit score estimation was not accurate enough. It acts as a preventive measure to dissuade borrowers from defaulting, as their previous contributions are affected if they cause a financial loss to the protocol. The punishment may result in the cancellation of all previous contributions, treating the borrowers as new users. Our approach carries the potential of losing some earnings and, consequently, incurring a loss. Moreover, the advantages require multiple loans to build up to a significant amount, but we believe that it will also bring stability.
\subsection{Mathematical model}

\begin{table}[t!]
\centering
\caption{Summary of main notations}
\begin{tabular}{|cl|}  
\hline
Symbol & Meaning \\ 
\hline 
$\alpha_i$ & interest rate at the bank at time $i$\\
$\beta$ & default probability for a borrower\\ 
$\gamma_i$ & loan amount at time $i$\\
$\delta_i$ & the interest rate for the customer at time $i$\\
$\Delta_{coll}$ & fraction of collateral reduced for a loan\\ 
$\rho$ & the margin of profit for the lender\\
$k\$$ & value of 1000 USD\\
\hline
\end{tabular}
\label{tableNotations}
\end{table}

We now offer the mathematical approach of our protocol to ensure profit based on the credit score estimation. The profitability model is based on the assumption that funds can either be placed at the bank (secure investment opportunity mentioned earlier) or put into a lending platform. Thus, the model keeps on a higher expected value of profit in the lending platform. The main notations are summarized in Table \ref{tableNotations}.

\noindent
Assuming the interest rate at the bank is $\alpha$, the default probability for a loan is $\beta$, and $\delta$ is the interest rate for the customer, then the total incentive to lend to a customer requires at least the following:

\begin{equation} 
\alpha  < (1-\beta)\cdot\delta
\label{eqn:StakingAndInterest} 
\end{equation}

 A borrower will be incentivized to offer a loan only if the interest rate to the customer is bigger than that at the bank, and big enough to cover losses.

 If the formula holds for a single default rate across multiple interest rates, then we arrive at the following formula:

\begin{equation} 
 \sum_{i=1}^{n} \alpha_i \cdot  \gamma_i < \sum_{i=1}^{n-1} \delta_i \cdot \gamma_i + (1-\beta) \cdot \gamma_n \cdot \delta_n 
\label{eqn:MultipleDefaultRates} 
\end{equation}

 This expresses that the expected value of gains loaning is bigger than the gains at the bank, though is dependent on knowing exactly the default rate. The right side is the expected gains we make on the client while accounting for the probability of default. First, we take into account how much money we have made until the present: the interest rate for each past loan and its value. The second element is the amount of the expected value of this loan. The left side reflects the other choice. It is the gain made at the bank (with different rates over time) if we chose each time to invest at the bank instead of lending.

We update the formula to reduce collateral by risking some of the past gains for a given user. To be profitable, we require:
  
\begin{equation} 
\sum_{i=1}^{n} \alpha_i \cdot  \gamma_i < \sum_{i=1}^{n-1} \delta_i \cdot \gamma_i   + (1-\beta) \cdot \gamma_n \cdot \delta_n - \beta \cdot \Delta_{coll} \cdot \gamma_n
\label{eqn:PartialRisk} 
\end{equation}

while also maintaining
\begin{equation} 
\Delta_{coll} \cdot \gamma_n \cdot \delta_n < Min \{ \gamma_n , 0.5 \cdot  \sum_{i=1}^{n-1} \delta_i \cdot \gamma_i \}
\label{eqn:RiskLower50}
\end{equation}

to ensure we never risk more than 50\% of previous gains or the loan's value.

As detailed earlier in this section, we risk some of the collateral gained in the past for a given user. Therefore, the loan can end in one of the following scenarios:

\emph{Case 1}: The borrower repays the loan fully with interest.  
The collateral is unlocked and returned to the borrower. This situation happens with a probability of $1-\beta$. The gains on loan will then be $\gamma_n \cdot \delta_n$.

\emph{Case 2}: The borrower fails to pay at the due date of repayment, thereby defaulting, or the value of the collateral passes under the threshold of liquidation. The protocol liquidates the collateral and the loan is terminated. This situation happens with a probability of $\beta$. The losses on loan will be $\Delta_{coll} \cdot \gamma_n$.

Based on \eqref{eqn:PartialRisk}, we require an interest rate bigger than that of the bank to ensure profit. We will add a margin $\rho$, reflecting how much we improve compared to the gains at the bank. Our condition will hence be to maintain the equation for every loan we authorize:
\begin{equation} 
\sum_{i=1}^{n} \alpha_i \cdot  \gamma_i \cdot (1+\rho) = \sum_{i=1}^{n-1} \delta_i \cdot \gamma_i   + (1-\beta) \cdot \gamma_n \cdot \delta_n - \beta \cdot \Delta_{coll} \cdot \gamma_n
\label{eqn:InterestBasedOnCreditScore} 
\end{equation}

All parameters are known at the request of the loan, aside from $\delta_n$ and $\Delta_{coll}$. We believe the interest rate should be close to the interest at the bank to be attractive to the user, though we plot an example of the relationship of those factors.

\notinconf{
\begin{figure*}[h]
    \centering
    \begin{tikzpicture}[node distance={50mm}, thick, main/.style = {draw, circle, minimum size=3cm}]
    \tikzset{edge/.style = {->}}
\node[] (Title1) at (-2,1.8) {Start of first loan :};
\node[main] (F1) {Alice};
\node[main] (F2) [right of=F1] {smart contract};
\draw[edge]  (3.9,-1) -- node[midway, below, sloped] {Loan money: 100\$} (1.1,-1);
\draw[edge]  (1.1,1) -- node[midway, above, sloped] {1BTC worth 150\$} (3.9,1);
\node[] (collateral_title) at (2.4,0.8) {Collateral};
\node[] (Title2) at (-2,-1.8) {End of first loan:};
\node[main] (F3) at (0,-4) {Alice};
\node[main] (F4) [right of=F3] {smart contract};
\draw[edge]  (3.9,-5) -- node[midway, below, sloped] {1BTC worth 150\$} (1.1,-5);
\draw[edge]  (1.1,-3) -- node[midway, above, sloped] {Loan repaid:} (3.9,-3);
\node[] (collateral_title) at (2.4,-3.3) {100\$+10\$};
\node[] (Title2) at (-2,-5.8) {Start of second loan:};
\node[main] (F5) at (0,-8) {Alice};
\node[main] (F6) [right of=F5] {smart contract};
\node[] (collateral_title) at (5,-8.5) {(gained 10\$}; 
\node[] (collateral_title) at (5,-9) {from Alice)};
\draw[edge]  (3.9,-9) -- node[midway, below, sloped] {Loan money:100\$} (1.1,-9);
\draw[edge]  (1.1,-7) -- node[midway, above, sloped] {0.95BTC worth 142.5\$} (3.9,-7);
\node[] (Title2) at (-2,-9.8) {End of second loan:};
\node[main] (F7) at (0,-12) {Alice};
\node[main] (F8) [right of=F7] {smart contract};
\draw[edge]  (3.9,-13) -- node[midway, below, sloped] {0.95BTC worth 142.5\$} (1.1,-13);
\draw[edge]  (1.1,-11) -- node[midway, above, sloped] {Loan repaid:} (3.9,-11);
\node[] (collateral_title) at (2.4,-11.3) {100\$+10\$};
\end{tikzpicture}
\caption{ The protocol to reduce the collateral needed.}
    \label{fig:first_braess}
\end{figure*}
} 

\section{Estimating the Default Risk}
\label{section_risk}
We now describe guidelines for the design of the credit score and how to evaluate its performance. We will therefore present here the directions we wish to explore and a proposed methodology.
\subsection{Credit score}
Due to scarcity of data, where the time window only spans 4 months, instead of looking at the probability of not paying down the whole debt over a specified period (say, a month), we look at the proxy of not paying down $50\%$ of the debt over the next two weeks. To understand what contributes to this probability, we want to know one's payment history, as one who pays more often and who pays significant amounts each time is less risky. Relatedly, we want to know one's activity in general on the platform. One's recent collateral to debt may also play a role, as one who has significant collateral with respect to one's debt demonstrates his liquidity position. Finally, we integrate an age of the account in the process to slow malicious parties and treat recently created accounts as suspicious. We borrow some concepts from~\cite{KHANDANI20102767} for constructing features.

To incorporate the points we recommend taking into account in the score, we take the following features:

\begin{itemize}
\item age of the first transaction account on the Ethereum network compared to the current transaction time.
\item number of transactions of the user with the contract over the last two weeks
\item number of payments in USDC of a debt to the protocol per user in the past two weeks
\item average of daily collateral to the debt over the last two weeks


\end{itemize}
\label{eqn:creditScore1} 

To keep things simple, we use a Probit classification model to predict the probability of not paying down $50\%$ of the existent debt in the next two weeks. Probit and logistic models are quick to implement, easily interpretable, and less prone to overfitting, in contrast to other machine learning models used, but suffer from the drawback of constructing linear boundaries. For the sake of this paper, we take advantage of the simplicity of this model. Probit and logistic models are very similar, with a slight variation in terms of the underlying distribution. We consider a Probit model, due to its ease of relating the impact of a feature directly on the probability, the target variable.

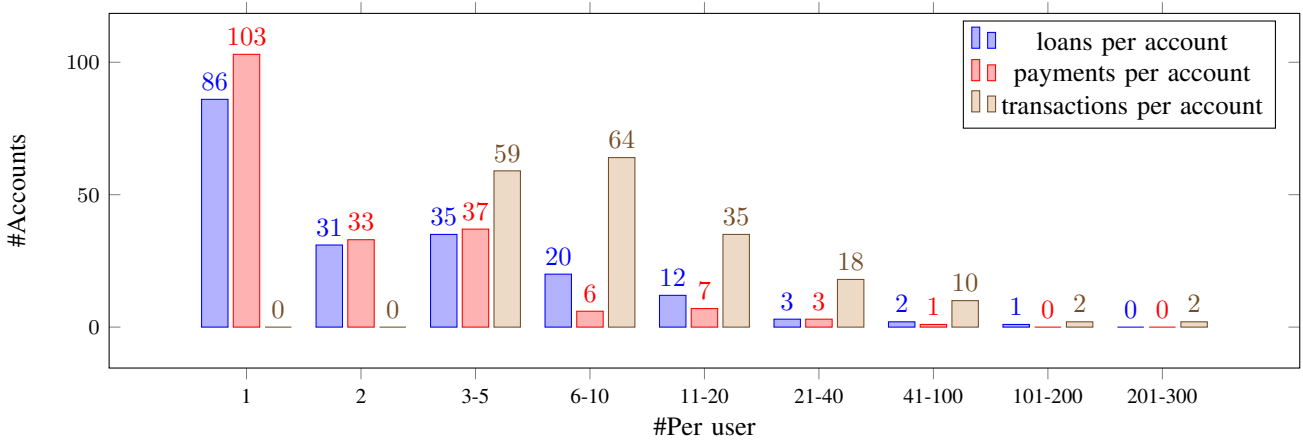
\begin{figure*}[t!]
\begin{tikzpicture}  
  
\begin{axis}  
[  
    ybar,  
    enlargelimits=0.15,  
    ylabel={\#Accounts}, 
    xlabel={\#Per user},  
    tick label style={font=\footnotesize},
    symbolic x coords={1, 2, 3-5, 6-10, 11-20, 21-40, 41-100, 101-200, 201-300}, 
    xtick=data,  
    width = 0.96 \textwidth,
    height=63mm,
     nodes near coords, 
    nodes near coords align={vertical},  
    ]  
\addplot coordinates {(1,86) (2,31) (3-5,35) (6-10,20) (11-20,12) (21-40,3) (41-100,2) (101-200,1) (201-300,0)};  

\addplot coordinates {(1,103) (2,33) (3-5,37) (6-10,6) (11-20,7) (21-40,3) (41-100,1) (101-200,0) (201-300,0)};

\addplot coordinates {(1,0) (2,0) (3-5,59) (6-10,64) (11-20,35) (21-40,18) (41-100,10) (101-200,2) (201-300,2)};

\legend{loans per account,  payments per account, transactions per account};

\end{axis}  
\end{tikzpicture}  
 \caption{Compound data analysis - Histogram of the number of interactions with the protocol per account based on the type of interaction: loans (withdraw value), payments (value return following previous loans), and transactions (all four types of interactions). There is a total of 190 accounts.}
\label{HistNumLoansUser}
\end{figure*}

\subsection{Risk estimation model}
We use a basic Probit model classifier to predict the probability of default in the next month. That is, we have
\begin{equation} 
Y = \Phi(\beta \cdot X+\epsilon),
\label{eqn:Probit} 
\end{equation} 
where $\Phi(\cdot)$ is the cumulative distribution function (CDF) of the Normal distribution, $Y$ is the probability of not paying down $50\%$ of the debt in the next two weeks, $X$ is the matrix of features discussed above, $\beta$ is the coefficient associated with $X=[X_1 \ldots X_n]$, for $n$ features, and $\epsilon$ is a Gaussian error term.

To understand the impact of each feature, we compute simple marginal effects:
\begin{equation} 
\frac{\partial Y}{\partial X_i} = \beta_i \phi(\beta \cdot X+\epsilon),
\label{eqn:MarginalEffects} 
\end{equation} 
where $\phi(\cdot)$ is the probability density function (PDF) of the Normal distribution. Given that the effect for a given $X_i$ can change, due to non-linearity, we take the average effect over all points in the data, also known as the Average Marginal Effect (AME).

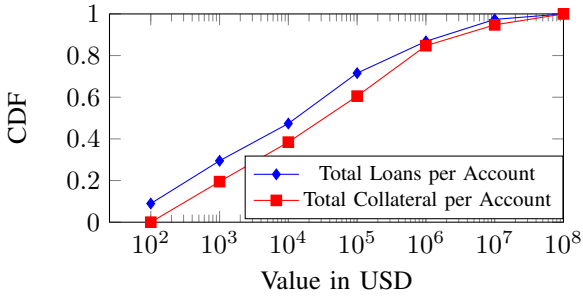
\begin{figure}
\begin{tikzpicture}
\begin{axis}
[ 
    xlabel={Value in USD},
    ylabel={CDF}, 
    xmax=1e8,
    xmode=log,
    ymin=0, ymax=1,
    xtick={0,1e2,1e3,1e4,1e5,1e6,1e7,1e8},
    height = 0.24\textwidth,
    width = 0.42\textwidth, 
    legend style={nodes={scale=0.8}, at={(1,0.32)}}
] 

\addplot[color=blue, mark=diamond*] coordinates {
(100,0.08947368421052632) (1000,0.29473684210526313) (10000.0,0.47368421052631576) (100000.0,0.7157894736842105) (1000000.0,0.868421052631579) (10000000.0,0.9736842105263158) (100000000.0,1.0)
};
\addlegendentry{Total Loans per Account};

\addplot[color=red, mark=square*] coordinates {  
(100,0.0) (1000,0.19473684210526315) (10000.0,0.38421052631578945) (100000.0,0.6052631578947368) (1000000.0,0.8473684210526315) (10000000.0,0.9473684210526315) (100000000.0,1.0)
};
\addlegendentry{Total Collateral per Account};

\end{axis}
\end{tikzpicture}
\caption{Loan and Collateral aggregated value per account}
\label{LoanCollAggregatedValue}
\end{figure}

\subsection{Additional Points to Consider}
Anonymity does present an additional risk compared to bank-based loans but we argue that the risk of default is already accounted for by the model. In fact, one of the model's strengths is that despite identity-related information being masked, it can still detect riskiness from other factors. Moreover, although anonymity can incentivize the borrower to default on their loan because their reputation is not on the line - given that they can just open another account and there is no legal recourse (similar to the case of sovereign debt, where reputation alone is not enough because there is no legal recourse \cite{bulow1988sovereign}) - the protocol sets in place levels of collateral and interest rate, based on the model's credit assessment, so that the lender is protected from default, and the collateral can easily be liquidated. Given that the collateral is a carefully chosen cryptocurrency, it can be liquidated by posting a transaction on the network.

\section {Analysis of Real Data of the Compound Lending System}
\label{section_compound_data}
We present here how the data was extracted and used for the loans pool.

\subsection{Overview on the data}
We extracted data concerning a Compound pool called USDC/Ethereum \cite{CompoundPoolUSDC} that emits CUSDCV3 for liquidity providers. The data was pulled using Etherscan API to get the transactions involved in the process and their receipts in order to pull the logs and their associated data associated. The data was pulled from the contract creation data (2022-08-13 05:35:17) up to (2022-12-11 13:35:11). The data from the contract Bulker (a helper contract to help users pack multiple transactions into one) was also pulled from (2022-08-24 20:00:59) up to (2022-12-12 20:48:35).

We observe a total of: 
\begin{itemize}
    \item 190 different borrowers
    \item 2975 transactions between the borrowers and the protocol, including:
    \subitem - 950 loans taken (every act where USDC is transferred from the protocol for the user with a negative balance)
    \subitem - 1152 withdraw collateral actions \subitem - 610 reimbursements actions.
\end{itemize}
The total value loaned is 154.6 Million USD. 
The total value of collateral put on the platform is 314.9 Million USD. 

The platform of Compound allows users to send USDC to the contract, and in exchange, when their balance is positive, they see their balance increase by an Earn APR. When the balance is negative (providing that they put collateral) their balance decrease by a Borrow APR. We consider being part of the loan only the money that the user did not provide to the platform before (i.e making sure her balance is negative). 

\textbf{Format of the data.}
To collect the data, we refer to several types of events (transaction types) as follows. 
\begin{itemize}
\item \emph{Supply Collateral} a user providing an authorized collateral token to the protocol. The supported collateral tokens are WETH, WBTC, Compound, Uniswap, ChainLink.
\item \emph{Supply USDC} - a user providing USDC to the protocol. This happens in two possible cases: (i) providing liquidity to the platform for potential future loans of other accounts; (ii) increasing the balance of an account as a return of an existing loan.
\item \emph{Withdraw USDC} -  a user pulling USDC from the protocol as part of a loan or in the case of a positive account balance.
\item \emph{Withdraw Collateral} - a user pulling existing collateral that he provided earlier. The event is allowed if the minimal value of collateral is maintained according to the existing loans of the account. 
\end{itemize}

\begin{figure}
\begin{tikzpicture}
\begin{axis}
[ 
    xlabel={Value (USD)},
    ylabel={CDF}, 
    xmax=1e8,
    xmode=log,
    ymin=0, ymax=1.0, 
    ymajorgrids,
    xtick={0,1e2,1e3,1e4,1e5,1e6,1e7,1e8},
    height = 0.24\textwidth,
    width = 0.42\textwidth, 
    legend style={nodes={scale=0.8}, at={(1,0.16)}}
] 

\addplot[color=blue, mark=diamond*] coordinates {
(100,0.08947368421052632) (1000,0.29473684210526313) (10000.0,0.48947368421052634) (100000.0,0.7473684210526316) (1000000.0,0.8842105263157894) (10000000.0,0.9947368421052631) (100000000.0,1.0)
};
\addlegendentry{Max Debt for each user at any time};

\end{axis}
\end{tikzpicture}
\caption{Maximum Loan debt value per account}
\label{MaxDebtUser}
\end{figure}

\begin{figure}
\begin{tikzpicture}
\begin{axis}
[ 
    xlabel={ratio},
    ylabel={CDF}, 
    xmin=1, xmax=7,
    ymin=0, ymax=1.0, 
    ymajorgrids,
    xtick={0, 1,1.5, 2, 3, 4, 6, 13,19},
    height = 0.24\textwidth,
    width = 0.42\textwidth, 
    legend style={nodes={scale=0.8}, at={(1.05,0.3)}}
] 
\addplot[color=black, mark=triangle]
coordinates{(1,0.0) (1.2,0.006493506493506494) (1.4,0.03896103896103896) (1.6,0.12337662337662338) (1.8,0.2857142857142857) (2,0.4155844155844156) (3,0.8181818181818182) (5,0.948051948051948) (7,0.974025974025974) (19,1.0)};
\addlegendentry{Average of daily Collateral to debt};
\addplot[color=purple, mark=diamond*] coordinates {
(1,0.0) (1.2,0.015789473684210527) (1.4,0.14210526315789473) (1.6,0.3526315789473684) (1.8,0.5) (2,0.6157894736842106) (3,0.8894736842105263) (4,0.9578947368421052) (6,0.9736842105263158) (13,1.0)
};
\addlegendentry{Collateral to debt upon maximal debt};
\end{axis}
\end{tikzpicture}
\caption{Loan distribution value per account}
\label{CollateralToMaxDebtUser}
\end{figure}

\begin{figure}
\begin{tikzpicture}
\begin{axis}
[ 
    xlabel={Time elapsed in hours},
    ylabel={CDF}, 
    xmax=2610,
    xmode=log,
    ymin=0, ymax=1.0, 
    ymajorgrids,
    xtick={1,10,100,800,2600},
    xticklabels={1,10,100,800, 2600},
    height = 0.24\textwidth,
    width = 0.42\textwidth, 
    legend style={nodes={scale=0.8}, at={(0.63,0.8)}}
] 

\addplot[color=green, mark=diamond*] coordinates {
(0,0) (1,0.19473684210526315) (2,0.21052631578947367) (10,0.25263157894736843) (100,0.5105263157894737) (1817,1.0) (2514,1.0)
};
\addlegendentry{Time to first payment};

\addplot[color=orange, mark=square*] coordinates {  
(0,0.0) (1,0.15263157894736842) (2,0.16842105263157894) (10,0.20526315789473684) (100,0.3473684210526316) (1817,0.9631578947368421) (2514,1.0)
};
\addlegendentry{Time to last payment };

\addplot[color=green, mark=triangle*] coordinates {
};
\addlegendentry{};

\addplot[color=orange, mark=*] coordinates {
};
\addlegendentry{};
\end{axis}
\end{tikzpicture}
\caption{Time elapsed from the first Loan to first/last Payment}
\label{TimeFirstLastPayment}
\end{figure}
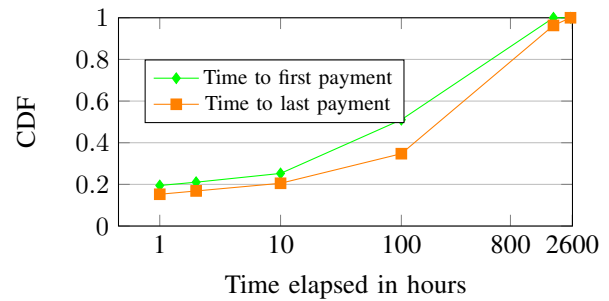


\subsection{Statistics and conclusions from the data}
Figure \ref{HistNumLoansUser} shows a histogram that reflects the number of users that took a defined number of loans. As shown, the majority of users have taken a single loan, and very few have over 10 loans. Moreover, the number of transactions in total with the protocol is relatively low,  ranging from 3 to 8 for half of the users, and up to 60 for 96\% of the users. It is then challenging to predict behavior based on this data alone.

Figure \ref{LoanCollAggregatedValue} presents the CDF of the total amount of loans taken  and the total amount of collateral provided (both in USD). The loans are between about 100\$ and up to 39M\$, with a median of around 15k\$. The collateral oscillates between 127\$ and 61M\$, with a median of around 31k\$. The majority of the users have total collateral of under 100k\$. 

Figure \ref{MaxDebtUser} shows the CDF of the maximal debt of an account during the evaluation period, ranging between 98\$ and up to 11M\$. Half of the users had a debt smaller than 15k\$. 

Figure \ref{CollateralToMaxDebtUser} shows how collateralized the maximum debt is  in the protocol for a user's largest debt. 60\% of users have a collateralization between 120\% and 200\%, leaving still an over-collateralization, for 40\% of users that oscillates between 200\% and up to almost 1300\%. The figures also show the average per-user daily debt to collateral ratio. Almost 80\% of users borrow under 50\% of the amount provided as collateral, making the borrow usage for most of them quite low.

\begin{figure}[t!]
 \centering

 

  \begin{tikzpicture}
    \begin{axis}[
    xmin=0, domain=0:110,
    xlabel={$\Delta_{coll}$ reduced loan value in \%},
    ylabel={$\delta_n$ interest rate in\%},
    ymin = 0, ymax = 100,
      ymajorgrids,
    legend style={at={(0.99,0.5)},anchor=west},
    height = 0.22 \textwidth, 
    width = 0.361  \textwidth]
        \addplot[color = purple] {100*(35.15+0.2*x)/80};
        \addplot[color = blue] {100*(15.05+0.2*x)/80};
        \addplot[color = red] {100*(10.025+0.2*x)/80};
        \addplot[color = green] {100*(6.005+0.2*x)/80};
    \legend{$\rho$=3\%, $\rho$=1\%,$\rho$=0.05\%, $\rho$=0.01\%}
    \end{axis}
\end{tikzpicture}

\caption{The required interest rate based on the risk of default, for interest rate $\alpha = 5\%$}
\label{requiredInterestRate}
\end{figure}
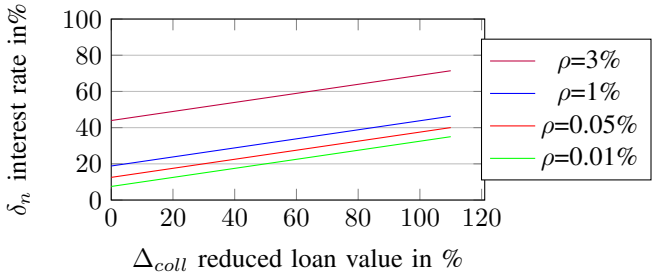

Figure \ref{TimeFirstLastPayment} shows the difference between the first loan and the first/last payment to the protocol. As we can see a small fraction of users pay back in an hour for the first time. We have a first payment that can be as long as 76 days after taking the loan, meaning the loan utilization time is relatively short. The same trends appear for the last payment.

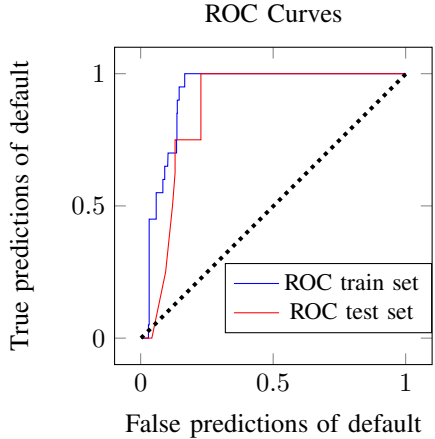
\begin{figure}
\centering
\begin{tikzpicture}
\begin{axis}[
  title={ROC  Curves},
  xlabel={False predictions of default},
  xtick = {0,0.5,1},
  ylabel={True predictions of default},
      height = 0.32\textwidth,
    width = 0.32\textwidth, 
    legend style={nodes={scale=0.9}, at={(1,0.32)}}
]

\addplot [blue] 
coordinates{ (0.0013966480446927375,0.0)(0.029329608938547486,0.0)(0.029329608938547486,0.05)(0.03212290502793296,0.05)(0.03212290502793296,0.45)(0.05865921787709497,0.45)(0.05865921787709497,0.55)(0.08379888268156424,0.55)(0.08379888268156424,0.6)(0.09078212290502793,0.6)(0.09078212290502793,0.65)(0.10195530726256984,0.65)(0.10335195530726257,0.7)(0.13547486033519554,0.7)(0.13547486033519554,0.75)(0.13687150837988826,0.75)(0.13687150837988826,0.85)(0.138268156424581,0.85)(0.138268156424581,0.9)(0.1452513966480447,0.9)(0.1452513966480447,0.95)(0.16620111731843576,0.95)(0.16620111731843576,1.0)(0.952513966480447,1.0)(1.0,1.0)};
\addlegendentry{ROC train set};
\addplot [red] 
coordinates{(0.00324675,0) (0.04220779,0) (0.04220779,0) (0.09415584,0.25) (0.09415584,0.25) (0.12012987,0.5) (0.12012987,0.5) (0.12987013,0.625) (0.12987013,0.75) (0.22727273,0.75) (0.22727273,1) (0.97077922,1) (1,1)};
\addlegendentry{ROC test set};


\draw[dotted, line width=1.5pt] (0,0) -- (1000,100);
\end{axis}
\label{ROCAUCCURVE}
\end{tikzpicture}
\caption{ROC (Receiver operating characteristic) Curves of the model predicting the probability of paying back half of the debt within two weeks}
\label{figure_roc}
\end{figure}

\section{Experimental Evaluation of the Protocol}
\label{section_evaluation}

We focus more on the impact of the features, rather than the performance of the model itself because we want to assess which features play important roles in the credit score.

\subsection{Illustration of the model trade-offs}
  Assuming a default risk of $\beta = 0.2$, a present loan of $\gamma_n = 100\$$ and past gains of $\sum_{i=1}^{n-1} \alpha_i \cdot  \gamma_i = $1000\$  at the bank and similar past gains in the lending platform $\sum_{i=1}^{n-1} \delta_i \cdot \gamma_i$= 1000\$. For $\rho=3\%$ and bank interest rate $\alpha = 5\%$, Equation~\ref{eqn:InterestBasedOnCreditScore} implies $(1000+0.05 \cdot 100)\cdot(1+0.03) = 1000 + (1-0.2)\cdot100\cdot\delta_n -0.2\cdot\Delta_{coll} \cdot 100$. Accordingly,  $35.15 = 80\cdot\delta_n -0.2 \cdot \Delta_{coll} \cdot 100$.
We illustrate that dependency  in Figure \ref{requiredInterestRate}. 
The figure shows, for a given interest at the bank, the interest rate of the loan and the portion of the collateral that is reduced as a trade-off between the two. Lowering collateral augments risk, and is compensated by a higher interest. This presents a heuristic of the solution to visualize the impact of different choices.

\begin{table}[t!]
\centering
\begin{tabular}{ c|cc }
\textbf{Feature} & Average Marginal Effect   \\
 &  (AME)   \\
\hline
Number of payments over past two weeks &  $-0.23064$ \\ 
\hline
Account Age & $-0.00497$ \\
\hline
Number of transactions over past two weeks & $0.00356$   \\
\hline
Collateral to the debt over the past two weeks & $0.00316$  \\ 
\end{tabular}
\caption{Feature impact on target variable}
\label{tab:impact}
\end{table}

\subsection{Evaluation of the risk estimation model}
As in~\cite{Score_Aave_Creditworthiness}, we use ROC (Receiver Operating Characteristic) as the performance metric, due to class imbalance. The imbalance is specifically about $3\%$ of accounts, that will not pay $50\%$ of the debt over the next 2 weeks. First, we do a simple $70/30\%$ train/test split, preserving the class imbalance. We examine the train dataset  and obtain ROC AUC of $92.3\%$. Then on test, we obtain an out-of-sample ROC AUC of $87.8\%$. This bodes well for the model's performance on a simple test set. Figure \ref{figure_roc} presents the ROC curves.



\subsection{Feature Impact}

For the whole dataset, we compute the AME for each feature. We present the contributions ranked (in absolute value) in Table \ref{tab:impact}. The most important feature in terms of its contribution towards paying down one's debt is the number of payments made, not the collateral-to-debt ratio. That is, we feel comfortable reducing one's collateral with our protocol because we can look at one's ability to make multiple payments recently to pay down one's debt, as that is a significant indicator of riskiness or lack thereof.



\section{Conclusions and Future Directions}
\label{section_conclusions}

In this paper  we presented a protocol for reducing the burden of collateral, while at the same time not detracting from the ability to assess risk. The presented model and the evaluation thereafter were basic, but sufficient to present contributions toward credit risk. To allow for more sophisticated modeling, as a next step, we must incorporate more features that have different look-back periods, as well as the gradient of one's liquidity position. Furthermore, we must allow for more out-of-sample testing to ensure the robustness of the model performance. Finally, to deal with the nature of the small dataset, we must test out Bayesian methods, like a Naive Bayes classifier, which typically performs better on smaller datasets. We should also contrast this with ensemble methods, such as gradient boosting and random forest algorithms.

\bibliographystyle{IEEEtran}
\bibliography{sample}

\end{document}